\documentclass[twocolumn,aps,preprintnumbers,longbibliography]{revtex4-2}

\usepackage[latin9]{inputenc}
\usepackage{amsmath}
\usepackage{amssymb}
\usepackage{graphicx}
\usepackage{esint}
\usepackage[scr=boondoxo, scrscaled=1.05]{mathalfa}
\usepackage{tikz-feynman}
\usepackage{xfrac}
\usepackage[hidelinks,bookmarks=false]{hyperref}
\allowdisplaybreaks

\makeatletter

\pdfpageheight\paperheight
\pdfpagewidth\paperwidth

\pdfoutput=1

\usepackage{epsfig}
\usepackage{graphics}

\makeatother

\begin{document}


\title{Angular spectrum of quantum  fluctuations in causal structure}
\author{\vspace{1pt}Craig Hogan} 
\affiliation{\vspace{1pt}University of Chicago, 5640 South Ellis Ave., Chicago, IL 60637}
\author{Ohkyung Kwon}
\affiliation{\vspace{1pt}University of Chicago, 5640 South Ellis Ave., Chicago, IL 60637}
\author{Nathaniel Selub}
\affiliation{\vspace{1pt}University of Chicago, 5640 South Ellis Ave., Chicago, IL 60637}
\begin{abstract}
\vspace{0pt}Scaling arguments are used to constrain the angular spectrum of distortions on boundaries of macroscopic causal diamonds, produced by Planck-scale vacuum fluctuations of causally-coherent quantum gravity. The small-angle spectrum of displacement is derived from a form of scale invariance: the variance and fluctuation rate of distortions normal to the surface of a causal diamond of radius $R$ at  transverse physical separation $c\tau\ll R$ should depend only on $\tau$, with a normalization set by the Planck time $t_P$, and should  not depend on $R$. For measurements on scale $R$, the principle leads to universal scaling for variance on angular scale $\Theta$, $\langle\delta\tau^2\rangle_\Theta\simeq\tau\:\!t_p\sim\Theta R\:\!t_P/c$\:\!, and angular power spectrum $C_\ell\sim (R\:\!l_P)/\ell^3$ at $\ell\gg1$. This spectrum is consistent with a relational model of holographic noise based on causally coherent virtual null gravitational shocks, a general picture conjectured for all $\ell$. The high $\ell$ scaling is contrasted with that predicted in some other quantum models, which differ by one power of angular wavenumber $\ell$ and are shown to predict excessive blurring of images from distant sources.\vspace*{1pt}
\end{abstract}
\maketitle

\section{Introduction}

There is no widely accepted theory of how quantum gravity works~\cite{HollandsWald2004,Stamp_2015}.
It is generally agreed that standard relativity and quantum mechanics become inconsistent below  the Planck scale $t_P=\sqrt{\hbar G/c^5}$,  determined by Planck's constant $\hbar$, Newton's constant $G$, and the speed of light $c$. However, there is no consensus on large-scale phenomenology of quantum geometry, apart from the requirement that any theory must agree with classical gravity on large scales. The standard quantization of linearized gravity, based on quantum field theory, is not causally coherent, and cannot form the foundation of a complete consistent theory, even on large scales~\cite{HollandsWald2004,Stamp_2015}.

A key breakthrough would be to identify a uniquely identifiable large-scale phenomenon associated with  quantum gravitational effects.
Such a research program  may be possible if quantum states of geometry are nonlocally coherent on null surfaces, which leads to observable correlations even on macroscopic scales. 
Predictions of causally coherent macroscopic effects from Planck scale fluctuations have been derived  from from semiclassical models of holographic noise\:\cite{Ng2000,Hogan_2017b,Hogan2017c,Mackewicz2021,Kwon2022}, and from  more  formal quantum  theories of causal fluctuations\:\cite{BanksZurek2021,Verlinde_2022,VZ,Zurek_2022,LiPRD}.

In these scenarios, macroscopic null surfaces  acquire  large-scale distortions due to Planck scale quantum fluctuations, with an amplitude much larger than the Planck time.  In principle, their correlations may be observable, either in the signals of suitable configured interferometers\:\cite{Ng2000,Hogan_2017b,Hogan2017c,Mackewicz2021,Kwon2022,holoshear,Richardson2020,VZ,Zurek_2022,LiPRD} or in the relic cosmic perturbations measured in the cosmic microwave background\:\cite{Hogan_2022,Hogan2023,hogan2023causal}.

In this paper we use scale invariance to make a model-independent estimate of the  small-scale angular spectrum of 
macroscopic corrugations of causal diamond surfaces and horizons due to coherent Planck scale  quantum gravitational vacuum fluctuations. 

\section{Angular  spectrum and correlation}

Consider distortions of the spherical boundary of a causal diamond of radius $R$  with displacement in time  $\delta \tau$, measured for example  as  displacements of clocks (as considered in e.g. ref.\;\cite{Mackewicz2021}). 
  In general the displacement is relational, specific to the vantage point of the  world line that defines the causal diamond.


We adopt  standard notation from the CMB literature (see e.g.~refs.\;\cite{Hogan_2022,Hogan2023}).
A function  on a unit sphere
with directions $\smash{\vec{\smash{\Omega}\vphantom{\scalebox{}[.96]{O}}}}$ labeled by standard polar coordinates
$ \theta,\:\!\phi$
is decomposed into spherical harmonics
$Y_{\ell m}$:
\begin{equation}\label{decompose}
\delta\tau(\vec\Omega)=
\sum_\ell \sum_m Y_{\ell m}(\theta,\phi) \, a_{\ell m}
\end{equation}
where $a_{\ell m}$ are the harmonic coefficients of the discrete 2D angular spectrum.
The  angular power spectrum  is related to the harmonic coefficients by
\begin{equation}\label{powerpiece}
C_\ell= \frac{1}{2\ell+1}
\sum_{m= -\ell}^{m=+\ell} | a_{\ell m}|^2.
\end{equation} 
The angular correlation function is defined as
\begin{equation}\label{CTheta}
    C(\Theta)= \langle 
    \delta\tau_1 \delta\tau_2
    \rangle_\Theta
\end{equation}
where the average is taken over all points on the sphere with angular separation 
$\Theta=|\vec\Omega_1-\vec\Omega_2|$.
It is related to the angular power spectrum by
 \begin{equation}\label{harmonicsum}
 C(\Theta) = \frac{1}{4\pi}\sum_\ell \:\! (2\ell +1) \;\! {C}_\ell \;\! P_\ell (\cos \Theta),
\end{equation}
where $P_\ell $  are the Legendre polynomials.

\section{Causally coherent noise}

Our main aim is to derive general scaling laws for $C_\ell$ and $C(\Theta)$ applicable to scaling of Planckian quantum noise in the high-$\ell$ limit. 
Extrapolation to low $\ell$  is  affected by causal symmetries, whose main features can be captured in a geometrical model of coherent distortions on intersecting spherical null surfaces, as described in the example in the Appendix.


We will consider  anisotropy generated in a broad class of  models based on the principle that  quantum gravitational vacuum fluctuation states are coherent in  causal diamonds. 
This general construction allows
states of  space-time  to have coherent causal structures consistent with  particles they couple to.
A theory that obeys the correspondence principle can have geometrical states that match the coherent  angular pattern 
 of distortions on a spherical null surface
produced by classical gravity.

Consider  quantum states of geometry that are coherent on causal diamonds, with holographic granularity at the Planck scale.
Vacuum fluctuations produce directional causal distortions with  nonlocal angular coherence on the surface of every causal diamond.
A causally-coherent model leads to vacuum fluctuations with  variance and coherence similar to  that generated by displacements by coherent gravitational shock waves that would be produced by a classical gas of many pointlike  photons with randomly oriented Planck scale momenta\:\cite{Hogan_2017,Hogan_2017b,Hogan2017c,Mackewicz2021,Kwon2022}, whose  number density  saturates gravitational bounds.  For  harmonic distortions that add in quadrature, 
the total variance of distortion on the surface of a causal  diamond of radius $R$ is of order~\cite{Mackewicz2021} 
\begin{equation}\label{macrovariance}
\langle  \delta\tau^2\rangle  \sim R \:\! t_P/c\:\!. 
\end{equation}
 This semiclassical estimate corresponds to the  total holographic noise in the emergence  of relational positions among world lines from a Planck scale quantum system with causal coherence, and agrees with the standard quantum limit  for such a system.
Approximately the same overall variance has been derived more formally from a
conformal field theory of near-horizon vacuum states\:\cite{BanksZurek2021,Verlinde_2022,VZ,Zurek_2022,LiPRD}.

  \section{Small-angle scaling  invariance}

In a scale-invariant theory, holographic distortions measured locally in a region of size $\ll R$  should  not depend on the scale $R$. 
This property imposes
  general constraints on
  how the variance in Eq.\:(\ref{macrovariance}) is distributed among different angular scales in a system of size $R$.

At very small angular separation, we can ignore the curvature of a causal diamond surface. A causal diamond that intersects a  small patch of the surface contributes the same physical fluctuation in the normal direction to any much larger diamond, independent of its size.
This property leads to the following form of scale-invariance  symmetry for the spectrum:  {\it on small angular scales,  correlations of  physical displacement 
corrugations $\delta\tau$ in the normal direction to a null surface at transverse spacelike separation $c\tau$, which fluctuate coherently on timescale $\tau$,  should not depend on the total radial propagation distance $R$, only on the transverse separation scale $c\tau$.}
With this kind of invariance, measured fluctuations only depend on the causal diamonds encompassed by an actual measurement, not on much larger systems it is embedded in.

Scale invariance means that the variance  of corrugation in patches of physical size $\tau$ matches the variance of causal diamonds of that size (Eq.\;\ref{macrovariance}):
\begin{equation}\label{scaling}
    \langle 
    \delta\tau^2
    \rangle_\tau \simeq  \tau \:\! t_P.
\end{equation}
The contribution to total variance from  patches of size  $\Theta\ll1$ is given by Eq.\:(\ref{CTheta}), so   scaling invariance 
in the small angle limit
\begin{equation}
    \ell\sim \pi/\Theta\sim \pi R/ c\:\!\tau\gg1
\end{equation} leads to variance in patches of size $\Theta$
\begin{equation}\label{varianceTheta}
C(0)- C(\Theta)\sim \langle 
    \delta\tau^2
    \rangle_\Theta\sim     \tau \:\! t_p\sim 
    \Theta R \:\! t_P/c\:\!,
\end{equation}
and an equivalent scaling of the power spectrum
\begin{equation}\label{spectralscaling}
    C_\ell  \sim C(<\Theta) /\ell^2 \sim
    \tau\:\!t_P /\ell^2\sim  \ell^{-3} R \:\! t_P/c\:\!. 
\end{equation}
With this scaling the bulk of the total anisotropy is always dominated by the large angular scales and low-order modes. It agrees with  scaling computed in concrete examples, for example,  the angular spectrum of gravitational shock waves from photons emitted in  directionally indeterminate $S$-decay in an EPR experiment\cite{Mackewicz2021}.

In general, the angular spectrum at low $\ell$, and the correlation function at large $\Theta$, have nontrivial structure that depart from these scalings, due to the curvature in null shock fronts not accounted for in this argument. A specific illustration of a complete causally coherent candidate model spectrum is presented in the Appendix.

\section{Small-angle spectrum in the geontropic model}
 The scale-invariant spectrum contrasts with that of the ``pixellon'' in the Verlinde-Zurek geontropic model\:\cite{VZ,Zurek_2022,LiPRD}, which yields a variance that grows at small angles: 
 \begin{equation}\label{VZvariance}
  \langle 
    \delta\tau^2
    \rangle_\Theta =    (R \:\! t_P/\sqrt{8\pi^3}c)
    \log(\pi/\Theta),
 \end{equation}
 and leads to a different high-$\ell$ scaling,
 \begin{equation}\label{VZspectrum}
 C_\ell\propto \ell^{-2}.
\end{equation}
 (See for example Eqs.\:(27) and (28) of ref.\;\cite{VZ}, Eq.\:(30) of ref.\;\cite{Zurek_2022}, or
 Eq.\:(57) of ref.\;\cite{LiPRD}.)
 
 The  geontropic model shows approximately the same magnitude for the macroscopic variance as the scale-invariant, causally coherent  holographic noise model just described, but 
 with contributions spread out slightly more evenly in angular scale. This spectrum is not consistent with our conjectured scale invariance, since the variation in a causal diamond increases with the size of other causal diamonds in which it is  embedded.

As seen in Eq.\:(\ref{VZvariance}), the VZ geontropic spectrum  leads to a variance
$\langle \delta\tau^2\rangle$ that decreases only logarithmically on scales less than $R$. 
With this spectrum, wavefront displacements  randomly vary across the aperture of size $A\sim \pi R/\ell$ by an amount 
\begin{equation}\label{aperture}
\langle c^2 \delta\tau^2 \rangle\sim (R \:\! l_P)/\log(R/A).
\end{equation}

This scaling  requires correlations of wavefront corrugations with a small transverse separation but large $R$ to ``remember'' that they have been travelling for a long time. We regard this as implausible because it violates the scale invariance described above.  
Unlike the spectrum derived from scale invariance, the variation in a patch depends on $R$ as well as $A$.

The difference in angular spectrum matters for experimental design. While  both spectra  have comparable total variance, the fluctuation power in the geontropic model is approximately evenly divided over logarithmic intervals of scale.  One consequence is that the signal fluctuation power predicted with the geontropic model in interferometer experiments is typically reduced by a factor \:\!$\sim$\,$1/ \log(R/l_P)$.

\section{Blurring of distant images}

The shallower spectrum in Eq.\:(\ref{VZspectrum}) leads to an observable physical effect, a blurring of images from distant astronomical sources (as noted in ref.\;\cite{Kwon2022}). A  distortion of causal structure should lead to a comparable distortion in the wavefronts of light propagating from distant sources. The tilt of a wavefront changes  the apparent angular position of a distant point source.
This effect of coherent geometrical distortions  is not the same as  previous models of  blurring based on local interactions of quantum space-time ``foam'' with propagating photons\:\cite{Perlman2015}.

Consider the displacement of wavefronts entering a telescope aperture $A$  from a source at distance $D$.  
The source appears in the image plane in a direction normal to the mean wavefront entering the aperture.
Fluctuations in mean wavefront tilt lead to an effect analogous to atmospheric seeing:   the apparent location of a source fluctuates on the sky, with an angular variance
\begin{equation}
    \langle\delta\Theta^2\rangle\sim c^2 \langle \delta\tau^2 \rangle_{\Theta\:\!=A\;\!\!/\:\!\!D}\, \big/ \;\!\! A^2
\end{equation}
 and a timescale $\tau\sim A/c$.
 In the image plane, a pointlike source appears to be a point jittering very fast in an angular patch of area  $\langle\delta\Theta^2\rangle$ around its expected location. As expected from scale invariance, the jitter does not depend on the distance to a source, only on the size of the causal diamonds determined by the size of the measurement apparatus, $A\sim c\tau$.

Let us estimate the blurring produced in the VZ geontropic model.  For a source with angular size distance $D\sim 1$\;\!Gpc   typical of  a source at high redshift, the variance of angular  position, which appears as a blurring in a time averaged image, is
\begin{equation}
 \langle\delta\Theta^2\rangle\sim  \frac{D c\:\!t_P}{\sqrt{8\pi^3} A^2} \log(D/A)\sim \Big(\frac{5\:\!\mu{\rm m}}{A}\Big)^{\;\!\!\!2} \:\!\Big(\frac{D}{1\:\!\rm Gpc}\Big).
\end{equation}
 The geontropic blurring does not depend on the wavelength of photons, only the aperture used to image them.
  Since standard  diffraction of a photon wave produces images of angular size \:\!$\sim$\,$\lambda/A$, geontropic blurring dominates  diffraction  at wavelengths shorter than few microns, for sources at cosmological distances. 

Distortions of this magnitude would have prevented high resolution imaging of cosmologically distant point sources at infrared and shorter wavelengths.
Contemporary astronomers routinely obtain subarcsecond  images 
($\theta< 5\times 10^{-6}$) of distant QSOs  with  apertures of the order of a meter, which   rules out  such blurring. To choose one  example,   the Hubble Space Telescope ($A=2.4\;\!{\rm m}$) makes diffraction-limited UV images of cosmologically distant objects as small as $\sqrt{\langle\delta\Theta^2\rangle}\sim 1.5\times 10^{-7}$, about ten  times smaller than the expected geontropic blurring.

The predicted blurring for the scale-invariant $\ell^{-3}$ spectrum  is  smaller by a very large factor, since the amount of  variation on every scale is determined by  causal diamonds of size $A$, not $D_H$:
\begin{equation}
 \langle\delta\Theta^2\rangle\sim   (A\:\!l_P)/A^2 \sim l_P/A\:\!,
\end{equation}
which produces a negligible effect on images. In this case, the wavefront tilt is of the same magnitude as the relational holographic displacement of mirrors in interferometers.\vspace{-2.5pt}

\section{Conclusion}
We have derived constraints of a general character on the angular structure of large-scale causal distortions from coherent quantum gravity. Different models of holographic coherent quantum gravity lead to different predictions for angular spectra, some of which are constrained by existing astronomical  data. These constraints complement inferences from primordial perturbations, and from direct interferometric experiments.

\newpage
\bibliography{angularspectrum}

\section*{Appendix: Spectral models with coherent null shocks}


\begin{figure*}
\begin{centering}
\includegraphics[width=\linewidth]{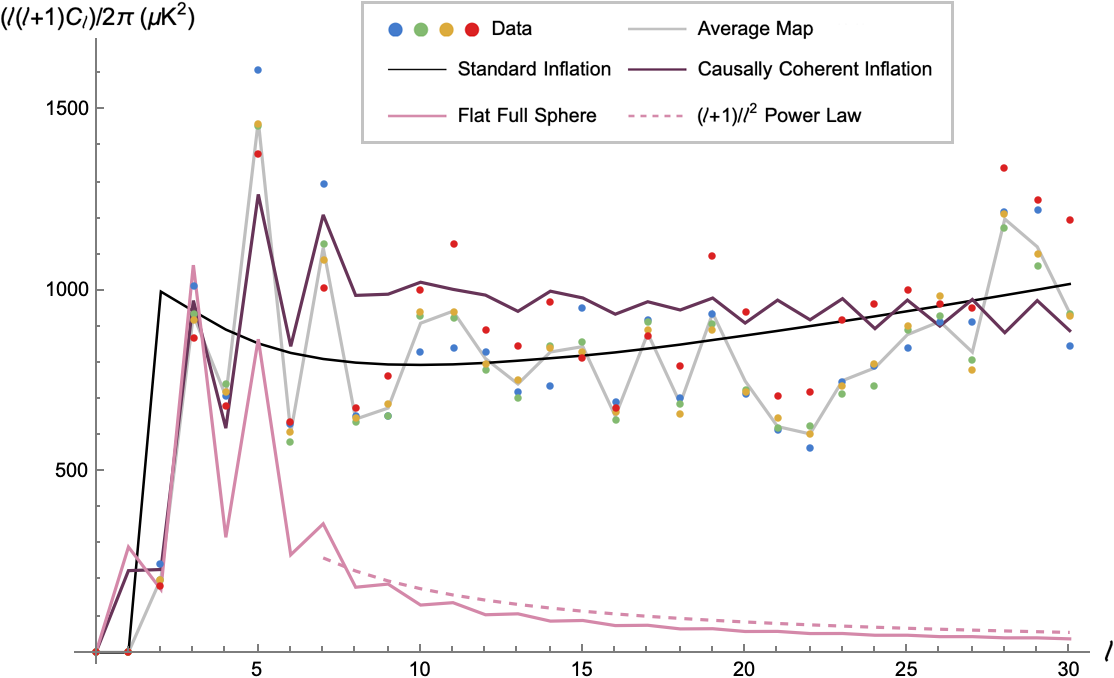}
\par\end{centering}
\protect\caption{Angular power spectra of causal distortions on spacelike spheres.  Causally coherent model spectra   are shown for   constant-time surfaces in slow-roll inflation  (from ref.\;\cite{Hogan2023}), and  in flat space-time (from formulae in the text). As a reference, the dashed curve approximately follows a $C_\ell \propto \ell^{-3}$ power law, which is the scale invariant limit at high $\ell$ in the flat case. For comparison, the standard  expectation for CMB temperature anisotropy is also shown, as well as  CMB data  for various Galaxy-subtracted maps and their average\:\cite{Hogan_2022,Hogan2023}. 
 \label{spectra}}
\end{figure*}

In recent previous work\:\cite{Hogan2023},  a classical noise model was developed
 to  analyze  holographic correlations of  large-angle Cosmic Microwave Background (CMB) anisotropy, 
 based on  virtual relational shock displacements at intersections of   inflationary horizons.
 It is instructive to adapt this approach to estimate  the large-angle spectrum of causal distortions on constant-time surfaces  in flat space-time.

In the model, a single coherent shock  produces coherent displacements on circular intersections of null spheres, described by
an axially symmetric kernel $\delta\tau(\theta)$ that depends on the polar angle $\theta$ from the shock axis.
A  noise realization  is a sum of $N\gg 1$  distortions from shocks in different  directions $\vec \Omega_i$:
\begin{equation}
     \delta \tau_{total}(\vec \Omega)= \sum_i^N \delta_i \, \delta\tau(\theta= |\vec \Omega- \vec \Omega_i|) /c\:\!,
\end{equation}
where $ \delta_i$ represents a random variable with zero mean and unit variance. For a model with holographic gravitational entropy,  $N\simeq (R/c\:\!t_P)^2$ and $\langle\delta\tau^2\rangle\sim t_P^2$, which leads to    distortions on macroscopic causal surfaces  given by Eq.\:(\ref{macrovariance}). Since the shocks  map onto a  duration $R/c$, the total variation corresponds to a Planck variance per Planck time, like a random walk~\cite{Hogan_2017,Hogan_2017b,Hogan2017c,Mackewicz2021,Kwon2022}.
For large $N$, the sum of many such shocks  produces  a universal 
 holographic angular power spectrum via Eq.\:(\ref{powerpiece}),   determined by the transform of the kernel:
\begin{equation}
 a_{\ell 0} = 2\pi \int_0^{\pi} d\theta \;\!\sin (\theta)\, Y_{\ell 0}^* (\theta)\,\delta \tau(\theta).
\end{equation}
For the CMB\:\cite{Hogan2023},  a  geometrical model for $\delta\tau(\theta)$ was constructed based on causal projections and axial displacements appropriate for slow-roll inflation (Fig.\;\ref{spectra}). The background curvature affects the high-$\ell$ spectrum, since classical displacements stretch along with the comoving background  outside the inflationary horizon.


\begin{figure}[h!]
  \centering
\vspace{2.5pt}\includegraphics[width=\linewidth]{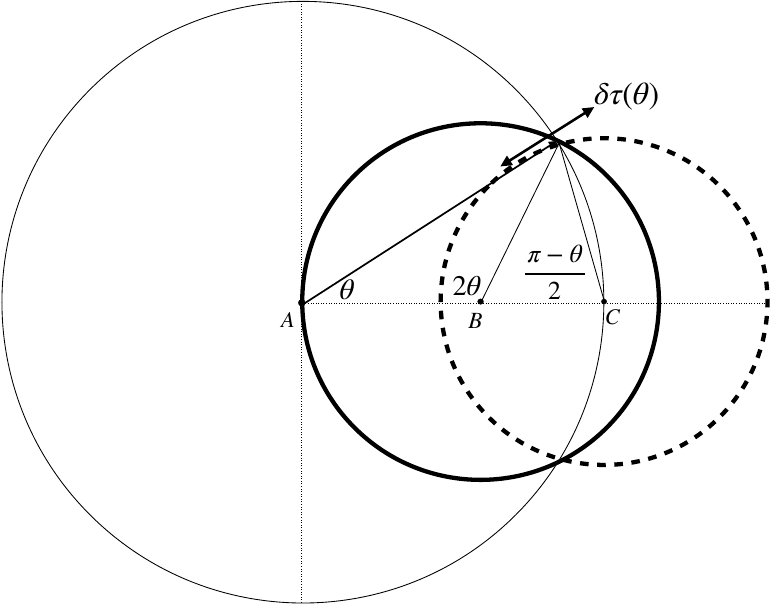}
  \caption{Azimuthal angles associated with intersections of $ABC$ spheres along an axis, which represent boundaries of causal diamonds entangled with shocks in the polar direction.\vspace{-12.5pt}
  \label{angles}}
\end{figure}

Flat space-time and inflation have the same conformal causal  structure, so  many elements of the model construction apply to both cases. 
For flat space-time,  we write the  kernel as 
\begin{equation}
    \delta\tau(\theta) \propto P_{ABC}(\theta)\,d_{AB}(\theta).
\end{equation}
The two factors represent  an
axial displacement $P_{ABC}$, and a  projection function $d_{AB}$,  for causal diamond surface intersections of world lines $A,B,C$  with angular relationships as shown in  Fig.\;\ref{angles}.

We adopt the same candidate form for $d_{AB}$ as the CMB model in ref.\;\cite{Hogan2023}, which has the same conformal causal structure:
\begin{equation}\label{simpleAB}
d_{AB}(\theta)\:\!= \;\!\cos (\theta)\:\!\cos(2\theta)\,
\Big[\;\!\!\cos(\theta)-\tfrac{1}{2} \sin(2\theta)\:\!\Big].
\end{equation}
This expression is an approximate interpolation that accounts for the coherent projections and displacements of causal shocks as viewed from the center of a causal diamond. 

 The $P_{ABC}$ term  in  flat space is not the same as inflation, but is constrained to agree with  the  invariant scaling derived above. The variance of  distortion for diamonds centered  on the surface of the distorted sphere is proportional to  duration, so the axial  displacement is
\begin{equation}
P_{ABC}(\theta)= 1/\sqrt{\sin(\theta/2)}.
\end{equation}

As shown in Fig.\;\ref{spectra}, the flat space-time spectrum reproduces the power law scaling expected at high $\ell$. At  $\ell\lesssim 7$, the model  resembles the spectrum measured on the CMB sky, with significant extra power in odd modes, especially $\ell=3,5,$ and 7. Both features are also consistent with the spectrum posited in a previous white paper, ref.\;\cite{Kwon2022}. These features  could be important in the design and interpretation of laboratory experiments.

\end{document}